\documentclass{astrobul}
\usepackage{graphicx}
\usepackage{natbib}
\usepackage{color}
\begin{document}

\journalinfo{2026}{81}{4}{1}[12]

\title{Physical Characteristics
of the Galaxy Defining the Oscillation Period of Librating Orbits
near the OLR of the Bar}

\author{A.~M. Melnik\address{1}\email{anna@sai.msu.ru} and
  E.~N.~Podzolkova\address{1},
  \addresstext{1}{Sternberg Astronomical Institute, Lomonosov Moscow
State University,  Moscow, 119234 Russia} }


\shortauthor{MELNIK and PODZOLKOVA}

\shorttitle{PHYSICAL CHARACTERISTICS OF THE GALAXY}

\submitted{June 15, 2026; accepted August 12, 2026}

\begin{abstract}
We studied librating orbits of stars near the Outer Lindblad
Resonance (OLR) of the bar using the angle-action variables of
Hamiltonian dynamics. The librating orbits change the direction of
their elongation relative to the major axis of the bar in a limited
range of angles. To calculate orbits, we used the model of the Galaxy
with an analytical bar which reproduces well the distributions of the
observed stellar velocities along the Galactocentric distance $R$
derived from the {\it Gaia} DR3 data. We compared the values of the
orbital oscillation period of a typical star obtained from the theory
and with the use of numerical simulation. Both values were found to
be in good agreement and are equal to 2.0 Gyr. It was shown that the
change in the average orbital radius, $\overline{R}$, of a star,
calculated over the time interval of one radial oscillation, is
regulated by the change in the direction of orbital elongation. As
the orbit is tilted relative to the minor axis of the bar in the
sense of Galactic rotation, the angular momentum of the star, $L$,
decreases, on average, causing a decrease in $\overline{R}$. As the
orbit is tilted in the opposite direction, $\overline{L}$ and
$\overline{R}$ increase. We derived a general formula for the
oscillation frequency, $\omega$, of orbits captured by libration near
the OLR of the bar. It turned out that the ratio of the  frequency of
orbit oscillations to the epicyclic frequency, $\omega/\kappa$, is
determined by the parameters $q_R$ and $\overline{p}$, where the
factor $q_R$ is the ratio of the amplitude of the radial force
generated by the bar to the average radial force produced by the
entire galaxy at the distance of the OLR, and the parameter
$\overline{p}$ characterizes the average elongation of orbits
captured by libration near the OLR.

\end{abstract}

\maketitle

\section{1. Introduction}

Modeling of stellar motion near resonances demonstrates the formation
of librating orbits that change the direction of their elongation
relative to the major axis of the bar. Furthermore, the orbital
oscillations are organized, they start predominantly in one direction
and have similar oscillation periods.

\citet{weinberg1994} developed a theory explaining the emergence of
librating orbits near resonances. He used a transition to the new
variables, namely, to the fast and slowly changing angles, where the
fast angle characterizes the star's position on the orbit while the
slow angle determines the direction of the orbit's elongation.
Averaging of  the Hamiltonian over the fast angle gives an
opportunity to represent  variations in the slow angle in the form of
oscillation equation.

The theory of the formation of librating orbits is presented in the
book by \citet[][chapters 3.5--3.7]{binney2008} and in the works by
\citet{monari2016, monari2017}. In recent years, the study of orbits
captured by resonances arouses great interest \citep{chiba2021,
chiba_schonrich2021, trick2021,al_kazwini2022}.

We explore a model of the Galaxy with an analytical bar
\citep{freeman1972,athanassoula1983,pfenniger1984, binney2008,
sellwood1993},  which forms an inner ring, r, located near the
corotation radius of the bar, and outer elliptical resonance rings
R$_1$ and R$_2$, located near the Outer Lindblad Resonance (OLR) of
the bar. Of the two outer rings, the ring R$_1$ is located closer to
the galactic center and is elongated perpendicular to the bar, while
the ring R$_2$ is located further from the center and is elongated
parallel to the bar \citep{schwarz1981, buta1991, byrd1994, buta1995,
buta1996, rautiainen1999, rautiainen2000, melnikrautiainen2009,
rautiainen2010, melnik2019}. The backbone of the resonance rings are
stable periodic orbits \citep{contopoulos1980, contopoulos1989}.

It turned out that librating orbits cause periodic changes in the
morphology of the outer rings, as well as the appearance of humps on
the profiles of the distributions of the radial velocity, $V_R$,
along the Galactocentric distance $R$ \citep{melnik2021, melnik2023}.
The median  oscillation period of librating orbits near the OLR of
the bar  is $\sim 2$ Gyr in our model \citep{melnik2024}. The study
of the stellar orbits in the model disk showed that a large group of
stars located in the so-called Hercules region are on librating
orbits \citep{melnik2025}.

The aim of this work is to compare the results of theoretical
calculations of stellar motions on librating orbits with results of
modeling and to obtain a formula connecting the most probable value
of the oscillation period of librating orbits near the OLR with the
physical characteristics of the Galaxy.

\section{2. Model and Observations}

We consider the model of the Galaxy that  reproduces well the
distributions of the radial, $V_R$, and azimuthal, $V_\varphi$,
stellar velocities along the Galactocentric distance $R$, derived
from the {\it Gaia} DR3 data \citep{prusti2016, katz2018, brown2021,
lindegren2021, vallenari2023}. In our previous work, we found that
the best agreement between the models and observations corresponds to
the bar's angular rotation velocity of $\Omega_b=55$ km s$^{-1}$
kpc$^{-1}$ and the Sun's position angle relative to the bar's major
axis of $\theta_\odot=-45^\circ$. In this model, the Corotation
Radius (CR) and the OLR radius of the bar are located at the
distances of $R_{\rm CR}=4.04$ and $R_{\rm OLR}=7.00$ kpc,
respectively. The bar turns on gradually, reaching the full strength
in four bar rotations, which corresponds to a bar growth time of
$T_g=0.45$ Gyr \citep[for more details, see ][]{melnik2021}.

Note that in galaxies with a bar, neither the angular momentum of a
star $L$ nor its total energy $E$ are conserved. However, if the bar
rotates with a constant angular velocity $\Omega_b$, then their
linear combination in the form of the Jacobi integral is conserved:

\begin{equation}
E_J= E-\Omega_b L
 \label{eq:E_J}
\end{equation}

\noindent \citep[for example,][formula 3.112]{binney2008}. In our
model, the value of $E_J$ at the OLR distance is conserved with a
relative accuracy of 10$^{-5}$.

In modeling  barred galaxies, the parameter $q_t(R)$ is often used,
which is the ratio of the amplitude of the azimuthal force generated
by the bar to the average  radial force created by the entire galaxy
at the same radius. The bar strength $Q_b$ is maximum value of
$q_t(R)$ along the radius \citep{sanders1980, combes1981,
athanassoula1983}. The bar strength $Q_b$ is used for classification
of bars. In our model,  maximum value of $q_t(R)$ corresponds to the
distance of $R=1.6$ kpc and amounts to $Q_b=0.3142$, which
corresponds to strong bars \citep{block2001, buta2004,
diaz-garcia2016}.

The distance to the Sun from the Galactic center is taken to be
$R_0=7.5$ kpc \citep[][]{glushkova1998, nikiforov2004,
eisenhauer2005, bica2006, nishiyama2006, feast2008, groenewegen2008,
reid2009b, dambis2013, francis2014, boehle2016, branham2017,
iwanek2023}. In general, choosing the value for $R_0$ in the range of
7--9 kpc has actually no effect on our results, since a shift in
$R_0$ causes a similar shift in the OLR radius.

\begin{figure}
\resizebox{8.0 cm}{!}{\includegraphics{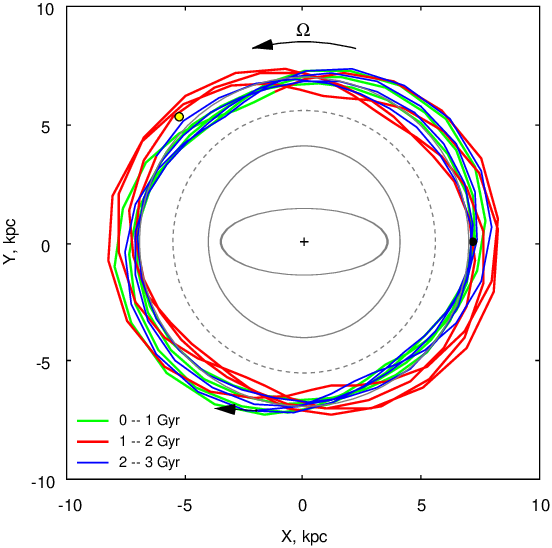}} \caption{Orbit of a
typical star on  librating orbit in the reference frame of the
rotating bar. The Galaxy rotates counterclockwise, but in the
reference frame of the bar, the chosen star rotates in the direction
opposite that of the Galactic rotation. The $X$ and $Y$ coordinates
are measured in the Galactic plane in the directions of the major and
minor axes of the bar, respectively. The initial coordinates and
velocities of the star are $R(0)=R_{\rm OLR}+0.125$ kpc,
$\varphi(0)=0$, $V_R=0$, and $V_\varphi=V_c-7=218$ km s$^{-1}$. The
orbital segments that the star passes during the time intervals of
0--1, 1--2, and 2--3 Gyr are shown in green, red, and blue,
respectively. The estimated position of the Sun and the initial
position of the star are shown as yellow and black circles,
respectively. It is clearly seen that during the time intervals of
0--1 and 2--3 Gyr, the stellar orbit is tilted to the right relative
to the minor axis of the bar, while during the time interval of 1--2
Gyr, it is tilted to the left.} \label{fig_orbit}
\end{figure}

\section{3. Theory}

For analytical studies, it is convenient to represent the galaxy's
potential as a sum of $\Phi=\Phi_0+\Phi_1$, where $\Phi_0$ is the
unperturbed, axisymmetric potential of the galaxy and
$\Phi_1(R,\varphi,t)$ is the perturbing potential, where
$\Phi_0\gg\Phi_1$. Suppose that the perturbing potential $\Phi_1$ can
be written as a product of two functions, one of which depends only
on the distance $R$, while the other -- only on the angle $\varphi$:

\begin{equation}
\Phi_1(R,\varphi,t)= \textrm{Re} \left[\alpha_b
\Phi_b(R)\mathrm{e}^{im (\varphi -\Omega_b t)}\right],
   \label{phi_1}
\end{equation}

\noindent where the potential $\Phi_b(R)$ is equal to the average
value of the potential created by the bar at the distance $R$, and
the coefficient $\alpha_b$ is  the ratio of the amplitude of the
radial force produced by the bar at the distance considered to the
average value of the radial force created by the bar at the same
distance:

\begin{equation}
 \alpha_b(R)= \frac{1}{2} \frac{\max(d{\Phi_b}/d{R}) -
 \min(d{\Phi_b}/d{R})}{<d{\Phi_b}/d{R}>}.
\label{eq:alpha_b}
\end{equation}

Note that the representation of the bar's potential as a product of
two functions depending on different variables (Eq.~\ref{phi_1})
describes well the radial-force distribution near the OLR, but
strongly overestimates the amplitude of the azimuthal force.
Generally, the separation of variables is a strong simplification of
the real picture.

Near the  resonances of the bar, the following condition must be
satisfied:

\begin{equation}
 l\kappa+m(\Omega-\Omega_b)=0
\label{eq:omega_s1}
\end{equation}

\noindent where $\kappa$ is the epicyclic frequency and $\Omega$ is
the angular velocity of the  rotation of a star in the azimuthal
direction. The coefficients $l=0$, $l=-1$, and $l=1$ correspond to
the resonances at the CR, ILR, and OLR distances, respectively.
Hereinafter, we will be interested in the potential perturbation near
the OLR ($l=1$). The main contribution to the perturbation produced
by the bar is made by the component $m=2$. Note that in our modeling,
we use the Ferrers bar, which also includes higher harmonics than
$m=2$. The $\alpha_b$ coefficient at the OLR distance takes the value
of $\alpha_b=0.0562$ in our model.

Let's consider the motion of a star in the epicyclic approximation.
The star moves along the epicycle with a constant angular velocity
$\kappa$, and the center of the epicycle during the time period
considered is located at the distance  $R_g$ and rotates with a
constant angular velocity  $\Omega_g$. In general, the use of the
epicyclic approximation is justified when the size of the epicycle is
much smaller than the distance to the Galactic center.

We  will use the angle-action variables ($\theta_R$, $J_R$) and
($\theta_\varphi$, $J_\varphi$). In the case of an axisymmetric disk,
the actions $J_R$ and $J_\varphi$ are integrals of  motion and do not
change with time. The action $J_\varphi$ is  the angular momentum of
the star $L$, and $J_R$ characterizes the energy of the epicyclic
motions.

The quantities ($J_R$, $J_\varphi$, $\theta_R$, $\theta_\varphi$) are
related to the coordinates and velocities in the cylindrical
coordinate system ($R$, $\varphi$, $V_R$, $V_\varphi$) by the
following relationships:

\begin{equation}
  J_R  =\frac{V_R^2}{2\kappa}+\frac{\kappa(R-R_g)^2}{2},
\end{equation}

\begin{equation}
J_\varphi=L=R V_\varphi,
\end{equation}

\begin{equation}
 \theta_R = \arctan(-\frac{V_R}{\kappa (R-R_g)}),
 \label{eq:theta_R}
\end{equation}

\begin{equation}
\theta_\varphi = \varphi+\Delta\varphi, \label{eq:theta_phi}
\end{equation}

\begin{equation}
\Delta \varphi = -\frac{\gamma}{R_g}\sqrt{\frac{2
J_R}{\kappa}}\sin\theta_R- \frac{J_R}{2} \frac{d \ln \kappa}{d
J_\varphi}\sin(2\theta_R), \label{eq:delta_phi}
\end{equation}

\noindent  where

\begin{equation}
\gamma =\frac{2\Omega}{\kappa}.
\end{equation}

Further, we will use only the first term in Eq.~\ref{eq:delta_phi},
which is  an order of magnitude larger than the second. In the case
of non-axisymmetric perturbations of the disk, the quantities $J_R$
and $J_\varphi$ cease to be integrals of motion.

Using Eqs.~\ref{eq:theta_phi} and \ref{eq:delta_phi} and retaining
only the first-order terms in $(R-R_g)$, we can represent the
perturbing potential (Eq.~\ref{phi_1}) near the distance $R_g$ as
follows:

$$
\Phi_1(R,\varphi,t)  = \textrm{Re} [\alpha_b \Phi_b(R) \mathrm{e}^ {i
m(\theta_\varphi-\Delta\varphi-\Omega_b t)}]
$$
$$
\approx \textrm{Re} \biggr( \alpha_b \Bigr[ \Phi_b(R_g)(1+i m \gamma
p \sin\theta_R) +
$$
\begin{equation}
-R_g \frac{d \Phi_b}{d{R}}(R_g)  p \cos\theta_R \Bigl] \mathrm{e}^{i
m (\theta_\varphi-\Omega_b t)} \biggl),
\end{equation}

\noindent where  the dimensionless quantity $p$

\begin{equation}
p(J_R, J_\varphi)  =\sqrt{2J_R/(\kappa R_g^2)}
\label{eq:p}\end{equation}

\noindent  characterizes the orbital elongation

\begin{equation}
R=R_g (1- p \cos\theta_R)
\end{equation}

\noindent \citep[][formula 3.261]{binney2008}, and is related to the
orbital eccentricity $e$ by the ratio $p \approx e^2/4$.

Note that the epicyclic frequency $\kappa$ depends on the angular
momentum of a star \citep[][formula 3.79a]{binney2008}:

\begin{equation}
\kappa^2=\frac{\partial^2{\Phi}}{\partial{R}^2}(R_g) +
\frac{3L^2}{R_g^4}. \label{eq:kappa}
\end{equation}

Let us  find the Fourier coefficients, $c_{j k}$, of the expansion of
the potential $\Phi_1$ in trigonometric harmonics near the distance
$R_g$, ignoring the rotation of the bar, i.~e. at the time $t=0$
\citep{weinberg1994, monari2016}:

$$
c_{j k}(J_R,J_\varphi)=\frac{1}{(2\pi)^2}
  \int_{0}^{2\pi} d\theta_R \int_{0}^{2\pi} d\theta_\varphi
$$
$$
\times \alpha_b \Phi_b(R)\mathrm{e}^{i m
(\theta_\varphi-\Delta\varphi)}\mathrm{e}^{-i
(j\theta_R+k\theta_\varphi)}
$$
$$
\approx  \delta_{k m}\frac{\alpha_b}{2\pi} \int_{0}^{2\pi} d\theta_R
\bigl[\Phi_b(R_g) + \frac{d \Phi_b}{d{R}}(R_g) (R-R_g)\bigr]
$$
$$
\times \bigl( 1+i m \gamma  p \sin \theta_R \bigr) \bigl(\cos
j\theta_R -i\sin j\theta_R \bigr)
$$
$$ \approx \delta_{k m} \alpha_b
\Bigl[ [\delta_{j 0}+\delta_{|j|1}\frac{m}{2} \textrm{sgn}(j)\gamma
p]\,\Phi_b(R_g)
$$
\begin{equation}
-\delta_{|j|1}  p \frac{R_g}{2}\frac{d \Phi_b}{d{R}}(R_g) \Bigr],
\label{eq:c_jk}
\end{equation}

\noindent where the Kronecker coefficients take the values
$\delta_{km}=1$ as $k=m$, and $\delta_{km}=0$ as $k \neq m$.

To study stellar motions near the bar's resonances, it is convenient
to transit to new variables using the generating function of the
second kind \citep{weinberg1994}:

\begin{equation}\label{eq:S}
 S=[l\theta_R+m (\theta_\varphi-\Omega_b t)] J_s+\theta_R J_f.
\end{equation}

\noindent The new angle-action variables ($\theta_s$, $\theta_f$,
$J_s$, $J_f$) are defined through the old variables ($\theta_R$,
$\theta_\varphi$, $J_R$, $J_\varphi$) as follows:

\begin{equation}
\theta_s  =l\theta_R+m (\theta_\varphi-\Omega_b t),
\label{eq:theta_s}
\end{equation}

\begin{equation}
J_\varphi=L=m J_s, \label{eq:J_phi}
\end{equation}

\begin{equation}
\theta_f = \theta_R, \label{eq:theta_f}
\end{equation}

\begin{equation}
J_R = l J_s+ J_f. \label{eq:J_R}
\end{equation}

The angle $\theta_s$ changes slowly and characterizes the orientation
of the stellar orbit relative to the major axis of the bar in the
reference frame of the rotating bar, while the angle $\theta_f$
changes rapidly and determines the motion of the star along its
orbit.

The motion of the star in  the new coordinates is described by the
Hamiltonian $H'$:

\begin{equation}
  H'(\theta_f,\theta_s, J_f, J_s)=H(\theta_R,\theta_\varphi, J_R, J_\varphi, t)+\frac{\partial{S}}{\partial{t}},
\end{equation}

\noindent where $\partial{S}/\partial{t}=-m \Omega_b J_s$. Therefore,

$$
H'=H_0+\Phi_1(\theta_f,\theta_s, J_f, J_s)-m\Omega_b J_s
$$

$$
=H_0+\textrm{Re} \Bigl(\sum_{j=-1}^1c_{jm}
\mathrm{e}^{i(j\theta_R+m\theta_\varphi)} \mathrm{e}^{-im \Omega_b
t}\Bigl) -m\Omega_b J_s
$$

\begin{equation}
 =H_0+\textrm{Re} \Bigl(\sum_{j=-1}^1c_{jm} \mathrm{e}^{i[(j-l)\theta_f+\theta_s]}\Bigl) -m\Omega_b J_s,
\label{eq:H'}
\end{equation}

\noindent where we used Eqs.~\ref{eq:theta_s} and \ref{eq:theta_f} to
express the angles $\theta_\varphi$ and $\theta_R$ in terms of
$\theta_s$ and $\theta_f$. Since the angle $\theta_f$ changes much
faster than the angle $\theta_s$, we can average the Hamiltonian $H'$
over the angle $\theta_f$, and the terms in the sum over $j$ vanish
if $j \neq l$. Therefore, the time-averaged Hamiltonian
$\overline{H'}$ can be written as:

\begin{equation}
 \overline{H'}=H_0+\textrm{Re} \Bigl(c_{lm}(J_f, J_s) \mathrm{e}^{i\theta_s}\Bigl) -m\Omega_b J_s.
\end{equation}

\noindent Then the equations of motion are:

\begin{equation}
  \dot{\theta}_s=
  \frac{\partial{\overline{H'}}}{\partial{J_s}}=\Omega_s + \textrm{Re}
  \Bigl(\frac{\partial{c_{lm}}}{\partial{J_s}} \mathrm{e}^{i\theta_s}
  \Bigl),
\label{eq:dot_theta_s}
\end{equation}

\begin{equation}
  \dot{J_s} =-\frac{\partial{\overline{H'}}}{\partial{\theta_s}}=
  -\textrm{Re}\Bigl(i c_{lm} \mathrm{e}^{i\theta_s} \Bigl).
\label{eq:dot_J_s}
\end{equation}

The resonance frequency $\Omega_s$ is determined from the equation:

\begin{equation}
  \Omega_s=\frac{\partial{(H_0-m\Omega_b J_s)}}{\partial{J_s}},
\label{eq:omega_s}
\end{equation}

\noindent where $H_0$ is the unperturbed Hamiltonian, and the linear
combination $(H_0-m\Omega_b J_s)$ is the Jacobi integral for the
unperturbed motion (Eqs.~\ref{eq:E_J},~\ref{eq:J_phi}). The quantity
$H_0$ can be represented as the sum of the energy of radial motions
$E_R$ and the energy of circular rotation $E_c$:

\begin{equation}
 H_0=E_R+E_c,
\label{eq:H_0}\end{equation}

\noindent where $E_R$ and $E_c$ can be written as follows:

\begin{equation}
 E_R=\kappa J_R,
\label{eq:E_R}
\end{equation}
\begin{equation}
 E_c=\Phi_0+ \frac{L^2}{2R_g^2},
\label{eq:E_c}
\end{equation}

\noindent \citep[][Formula 3.261]{binney2008}. Using
Eq.~\ref{eq:J_phi}, \ref{eq:J_R}, \ref{eq:omega_s}, \ref{eq:H_0},
\ref{eq:E_R}, and \ref{eq:E_c}, we can  easily show that the
resonance frequency $\Omega_s$ is:

\begin{equation}
\Omega_s= l\kappa+m(\Omega-\Omega_b). \label{eq:omega_s2}
\end{equation}

Differentiating Eq.~\ref{eq:dot_theta_s} and also using
Eq.~\ref{eq:dot_J_s} and neglecting terms of higher order than
$o(\Phi_1/\Phi_0)$, we obtain the oscillation equation:

\begin{equation}
\ddot{\theta}_s= \textrm{Re} \Bigl(i \, \Bigl[\,\Omega_s
\frac{\partial{c_{lm}}}{\partial{J_s}}
-\frac{\partial{\Omega_s}}{\partial{J_s}} c_{lm}  \Bigl]
\mathrm{e}^{i\theta_s} \Bigl). \label{eq:ddot_theta_s_1}
\end{equation}

\noindent The first term in parentheses is small because $\Omega_s
\approx 0$ near resonances, but the derivative
$\partial{\Omega_s}/\partial{J_s}$ can differ significantly from
zero. Therefore, Eq.~\ref{eq:ddot_theta_s_1} can be rewritten as
follows:

\begin{equation}
\ddot{\theta}_s= \Bigl(\frac{\partial{\Omega_s}}{\partial{J_s}}
c_{lm} \Bigl) \, \sin \theta_s. \label{eq:ddot_theta_s_2}
\end{equation}

\noindent In general, oscillations of the angle $\theta_s$ are not
harmonic and the oscillation frequency depends on the angle
$\theta_s$. The current value of the oscillation frequency $\omega$
for different values of the angle $\theta_s$ can be estimated as
follows:

\begin{equation}
\omega^2= -\frac{\partial{\Omega_s}}{\partial{J_s}}  c_{lm}
\frac{\sin \theta_s}{\theta_s},
 \label{eq:omega^2}
\end{equation}

\noindent and the current value of the oscillation period of the
angle $\theta_s$ is:

\begin{equation}
P=2\pi/\omega. \label{eq:per}
\end{equation}

The aim of this work is to compare the oscillation frequencies of the
angle $\theta_s$ derived from the theory and obtained with the use of
numerical simulation.

\begin{figure*}
\resizebox{\hsize}{!}{\includegraphics{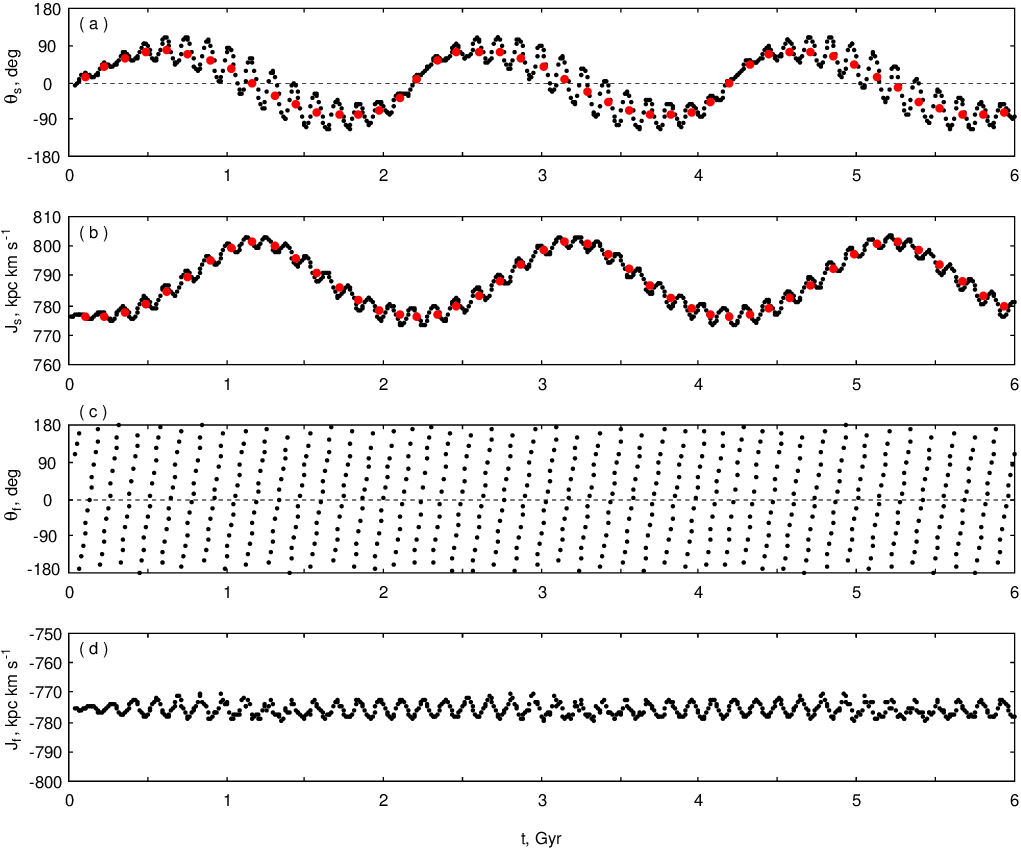}} \caption{
Variations of the variables $\theta_s$ (a), $J_s$ (b), $\theta_f$
(c), and $J_f$ (d), describing the motion of the chosen star
(Fig.~\ref{fig_orbit}), over time. Black dots correspond to the
instantaneous values of the variables every 10 Myr, and red dots
represent the values, averaged over the interval of one radial
oscillation. It is clearly seen that the oscillations of the slow
angle $\theta_s$ (a) and the action $J_s$ (b) have fast and slow
components with the periods of 0.13 and 2.0 Gyr, respectively.
Instantaneous values of the slow angle $\theta_s$ lie in the range
[$-115$, $115^\circ$], and its average values are in the range
$\overline{\theta_s} \in [-80, 80^\circ]$. The fast angle $\theta_f$
(c) increases almost uniformly with time. Oscillations of the action
$J_f$ (d) have only  a fast component.} \label{fig_ang_act}
\end{figure*}
\begin{figure*}
\resizebox{\hsize}{!}{\includegraphics{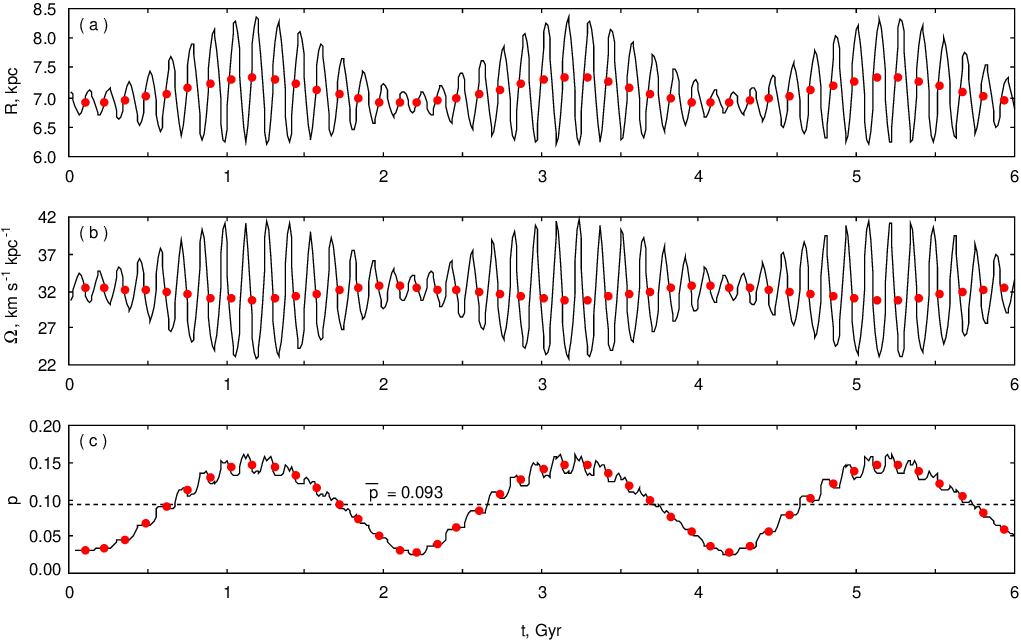}} \caption{ Changes
in the distance $R$ (a), angular velocity $\Omega=V_\varphi/R$ (b),
and orbital elongation  $p$ (c) of the  star under consideration over
time. Changes in the instantaneous values of the variables are shown
by the black line. The average values, $\overline{R}$,
$\overline{p}$, and $\overline{\Omega}$, calculated over the interval
of one radial oscillation, are shown by the red circles. The black
dashed line in panel (c) corresponds to the average value of the
orbital elongation $\overline{p}=0.093$. It is clearly seen that the
changes in $R$ and $\Omega$ have the form of beats and occur in
antiphase: the larger the average radius of the orbit $\overline{R}$,
the greater its elongation $\overline{p}$, and the smaller the
average angular velocity $\overline{\Omega}$.} \label{fig_r_omega_p}
\end{figure*}

\section{4. Results}

\subsection{4.1 Motion of a typical star}

Let us consider the motion of a typical star on a librating orbit
near the OLR of the bar. The initial distance of the star is $R(0) =
R_{\rm OLR} + 0.125$ kpc, and the initial velocities have their most
probable values at this distance, $V_R = 0$ and $V_\varphi = V_c - 7
= 218$ km s$^{-1}$, where $V_c$ is the velocity of the rotation
curve. The difference $-7$ km s$^{-1}$ between the rotation curve
velocity and the most probable value of the azimuthal velocity arises
due to the asymmetric drift  \citep[][section 4.8]{binney2008}. The
fact that the star in question is initially located not exactly at
the OLR radius, but slightly farther from the center, is due to the
fact that this distance corresponds to the most probable value of
$R(0)$ for librating orbits. The distribution of initial distances
and velocities of stars on librating orbits is described in detail in
\citep[][section 4.2]{melnik2024}.

Figure~\ref{fig_orbit} shows an orbit of a typical star on librating
orbit in the  reference frame of the rotating bar. The Galaxy rotates
counterclockwise. In the bar's reference frame, the star under
consideration  rotates in the direction opposite that of the Galactic
rotation. The orbital segments that the star passes over the time
intervals 0--1, 1--2, and 2--3 Gyr are shown in green, red, and blue,
respectively. It is clearly seen that during the time periods 0--1
and 2--3 Gyr, the stellar orbit is tilted to the right (i.~e.,
opposite that of Galactic rotation) relative to the minor axis of the
bar, and during the time period 1--2 Gyr, it is tilted to the left
(in the sense of Galactic rotation).

Figure~\ref{fig_ang_act} shows the variations in the angle-action
variables: $\theta_s$, $J_s$, and $\theta_f$, $J_f$, describing the
motion of the chosen star. Shown are the instantaneous values of the
variables (black dots) and the values averaged over the interval of
one epicyclic period or one radial oscillation of the star (red
circles). It is clearly seen that the variations in the slow angle
$\theta_s$ and the action $J_s$ have fast and slow components with
the periods of 0.13 and 2.0 Gyr, respectively. The oscillations of
$\theta_s$ and $J_s$ occur with a shift of $90^\circ$. Instantaneous
values of the slow angle, $\theta_s$, vary in the range [$-115$,
$115^\circ$], while its average values lie in the narrower range,
$\overline{\theta_s} \in [-80, 80^\circ]$. The fast angle $\theta_f$
increases almost linearly with time, and the values $\theta_f=\pm
180^\circ$ correspond to the same point on the epicycle. Oscillations
of the action $J_f$ have only  a fast component. Note that near the
OLR, the increase in the angle $\theta_R$ is mostly compensated by a
decrease in the angle $m (\theta_\varphi - \Omega_b t)$
(Eq.~\ref{eq:theta_s}).

Figure~\ref{fig_r_omega_p} shows the variations in the distance $R$
(a), angular velocity $\Omega=V_\varphi/R$ (b), and orbital
elongation $p$ (c) (Eq.~\ref{eq:p}) of the chosen star over time.
Changes in the instantaneous values of the variables are shown by the
black line. The average values, $\overline{R}$, $\overline{p}$, and
$\overline{\Omega}$, calculated over intervals of one radial
oscillation, are shown by red circles. It is clearly seen that the
changes in $R$ and $\Omega$ have the form of beats and occur in
antiphase. The larger the average orbital radius $\overline{R}$, the
greater the orbital elongation $\overline{p}$, and the smaller the
average angular velocity $\overline{\Omega}$.

We approximated the dependence of the bar potential $\Phi_b(R)$,
averaged over the azimuthal angle $\varphi$, on the distance $R$ by
the expression:

\begin{equation}
\Phi_b(R)= K \Bigl(\frac{a}{R}\Bigl)^k,
 \label{eq:a/R}
\end{equation}

\noindent where $a$ is the semi-major axis of the bar.  A similar
representation for the bar's potential was used  by
\citet{dehnen2000}. In our model, $a=3.5$ kpc \citep{melnik2021}. The
power $k$ was determined by searching through the range $k \in [0.1,
5.0]$ with a step of $\Delta k=0.1$. The constant $K$ was calculated
at each step using the least-squares method. We considered the
distance interval $a<R<11$ kpc, since we are interested in the bar's
potential in the OLR neighborhood. The upper limit of 11 kpc is due
to the fact that orbit integration stops as soon as a star moves
beyond 11 kpc. It turned out that the best agreement between
$\Phi_b(R)$ and the expression (\ref{eq:a/R}) corresponds to the
power $k=1.0$ and the coefficient $K=-14935 \pm9$ km$^2$ s$^{-2}$.

\begin{figure}
\resizebox{\hsize}{!}{\includegraphics{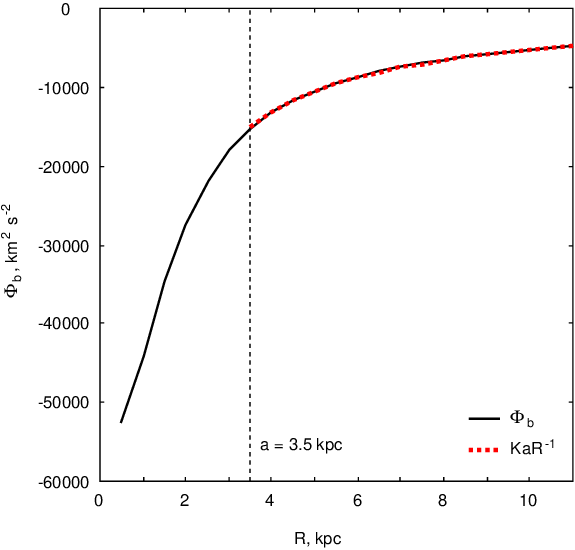}} \caption{
Dependence of the bar's potential $\Phi_b(R)$, averaged over the
azimuthal angle $\varphi$, on the distance $R$ (the black line).
Approximation of $\Phi_b(R)$ by the expression $\Phi_b(R)=K aR^{-1}$
over the distance range $R>a$, where $a=3.5$ kpc is the semi-major
axis of the bar (red dashed line).} \label{fig_pot}
\end{figure}

Figure~\ref{fig_pot} shows the dependence of the bar potential
$\Phi_b$, averaged over the azimuthal angle $\varphi$, on the
distance $R$, as well as its approximation by the expression $Ka/R$
over the distance interval $a<R<11$ kpc. We can see a good agreement
between the $\Phi_b(R)$ distribution and its approximation. The
standard deviation is $\sigma=263$ km$^2$ s$^{-2}$.

Hereinafter, we  assume that the average value of the Galactocentric
distance $\overline{R}$, calculated over the interval of one radial
oscillation of the star, corresponds to the position of the epicyclic
center:

\begin{equation}
R_g=\overline{R}.
 \label{eq:r_g}
\end{equation}

\noindent Note that the oscillations in the distance to the center of
epicycle $R_g$ (Fig.~\ref{fig_r_omega_p}a) and the oscillations of
the average angular momentum $J_s$ (Fig.~\ref{fig_ang_act}b) occur in
phase, i.~e., an increase in the average angular momentum $L$
($L=mJ_s$) is accompanied by an increase in $R_g$, and vice versa, a
decrease in $\overline{L}$ is accompanied by a decrease in $R_g$.

To calculate the angular oscillation frequency $\omega$ of   the
orbital elongation direction of the chosen star, we need to know the
values of the angle $\theta_s$, the derivative
$\partial{\Omega_s}/\partial{J_s}$, and the Fourier coefficient
$c_{12}$ (Eq.~\ref{eq:omega^2}).

The derivative $\partial{\Omega_s}/\partial{J_s}$ is determined by
the expression

\begin{equation}
\frac{\partial{\Omega_s}}{\partial{J_s}} = lm\frac{3L}{\kappa R_g^4}
+ \frac{m^2}{R_g^2},
 \label{eq:dir_omega_s}
\end{equation}

\noindent obtained with the use of Eqs.~\ref{eq:kappa},
\ref{eq:J_phi}, and \ref{eq:omega_s2}, as well as the relation
$\Omega=L/R_g^2$.

To calculate the Fourier coefficient $c_{12}$, we used
Eq.~\ref{eq:c_jk}, where $j=l=1$ and $k=m=2$. Note that the Fourier
coefficients depend on both the bar potential $\Phi_b$ and  its
derivative $d{\Phi_b}/d{R}$ (Fig.~\ref{fig_pot}). For the orbital
elongation $p$, we used the values of $\overline{p}$, averaged over
the interval of one radial oscillation. The $\gamma$ coefficient is
practically constant, lying in the range 1.403--1.408.

\begin{figure*}
\resizebox{\hsize}{!}{\includegraphics{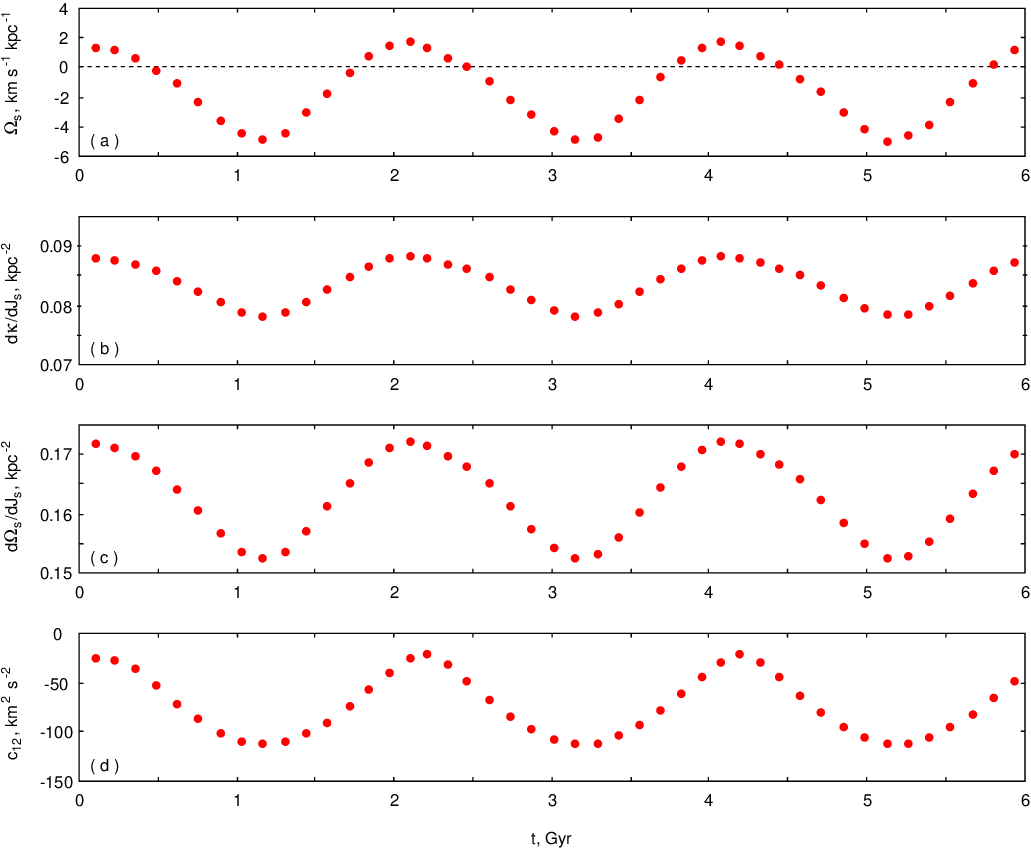}} \caption{
Variations in the average values of the resonance frequency
$\Omega_s$ (a), the derivatives $d{\kappa}/d{J_s}$ (b) and
$\partial{\Omega_s}/\partial{J_s}$ (c), and the Fourier coefficient
$c_{12}$ (d), calculated over the interval of one radial oscillation
of the star under consideration (red circles). It is clearly seen
that the resonance frequency $\Omega_s$ takes both positive and
negative values, the derivatives $d{\kappa}/d{J_s}$ and
$\partial{\Omega_s}/\partial{J_s}$ have only positive values, and the
Fourier coefficient $c_{12}$ takes only negative values. All listed
variables oscillate in phase.} \label{fig_omega_s}
\end{figure*}

Figure ~\ref{fig_omega_s} shows  variations in the average values of
the resonance frequency $\Omega_s$ (a), the derivatives
$d{\kappa}/d{J_s}$ (b) and $\partial{\Omega_s}/\partial{J_s}$ (c),
and the Fourier coefficient $c_{12}$ (d) (red circles), calculated
over the interval of one radial oscillation. It is clearly seen that
the resonance frequency $\Omega_s$ takes both positive and negative
values. The derivatives $d{\kappa}/d{J_s}$ and
$\partial{\Omega_s}/\partial{J_s}$ have only positive values, while
the Fourier coefficient $c_{12}$ takes only negative values. Note
that all  variables considered oscillate in phase.

Using the average values of the variables $\theta_s$,
$\partial{\Omega_s}/\partial{J_s}$, and $c_{12}$, we can calculate
the average values of the  frequency $\omega$ of  oscillations of the
orbital elongation direction (Eq.~\ref{eq:omega^2}).

\begin{figure*}
\resizebox{\hsize}{!}{\includegraphics{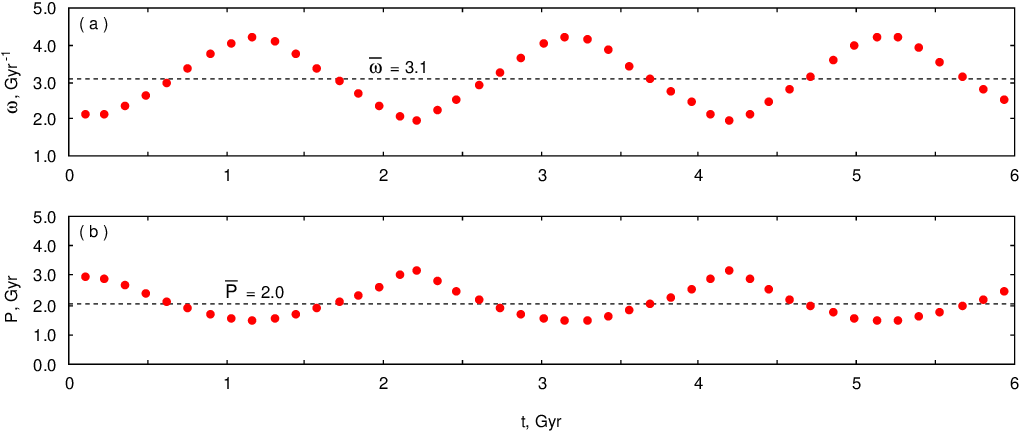}} \caption{Variation
in the frequency $\omega$ (a) and  the period $P$ (b) of oscillations
of the orbital elongation direction of the chosen star, calculated
with the use of Eqs.~\ref{eq:omega^2} and~\ref{eq:per} over the time
interval of one radial oscillation. The values of the frequency
$\omega$ and the period $P$, averaged over the time interval $t=0$--6
Gyr (the dashed lines), are $\overline{\omega}=3.1\pm 0.1$ Gyr$^{-1}$
and $\overline{P}=2.0\pm 0.1$ Gyr, respectively.} \label{fig_w_per}
\end{figure*}

Figure~\ref{fig_w_per} shows  variations in the frequency $\omega$
(a) and  oscillation period $P$ (b) of the orbital elongation
direction, calculated over the interval of one radial oscillation
(Eq.~\ref{eq:omega^2},~\ref{eq:per}). The value of the frequency
$\omega$, averaged over the time interval $t=0$--6 Gyr, is
$\overline{\omega}=3.1\pm 0.1$ Gyr$^{-1}$, and the corresponding
period amounts to  $\overline{P}=2.0\pm 0.1$ Gyr.

Thus, the average value of the oscillation period of  the orbital
elongation direction of a typical star, $2.0\pm0.1$ Gyr, calculated
with the use of analytical theory, coincides with the value obtained
by integration of the orbit.


\subsection{4.2 Reason for the appearance of librating orbits}

\begin{figure*}
\resizebox{\hsize}{!}{\includegraphics{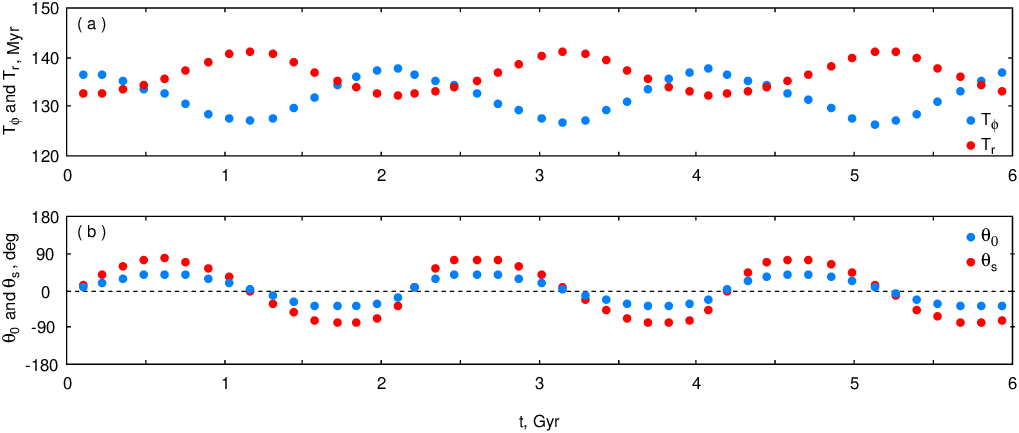}} \caption{(a)
Variations in the periods of the radial $T_r$ (red circles) and
azimuthal $T_\varphi$ (blue circles) oscillations of the chosen star.
The periods $T_r$ and $T_\varphi$ correspond to the frequencies
$\kappa$ and $2(\overline{\Omega}-\Omega_b)$, respectively. (b)
Variations in the slow angle $\overline{\theta_s}$ (red circles),
obtained from theory (Eq.~\ref{eq:theta_s}), and in the angle
$\theta_0$ (blue circles), obtained with the use of numerical
simulations. Both angles determine the orientation of the orbit
relative to the major axis of the bar over the interval of one radial
oscillation. It is seen that the angles $\overline{\theta_s}$ and
$\theta_0$ oscillate in phase, but the phase of the $\theta_s$
oscillations is determined up to a shift of $\pm \pi$. The amplitude
of the $\overline{\theta_s}$ oscillations is approximately twice as
large as the amplitude of variations in the $\theta_0$ angle. A
comparison of the dependences (a) and (b) shows that the increase
(decrease) in the $\theta_0$ angle occurs over the time intervals in
which $T_\varphi>T_r$ ($T_r>T_\varphi$).} \label{fig_t_phi_t_r}
\end{figure*}
\begin{figure*}
\resizebox{\hsize}{!}{\includegraphics{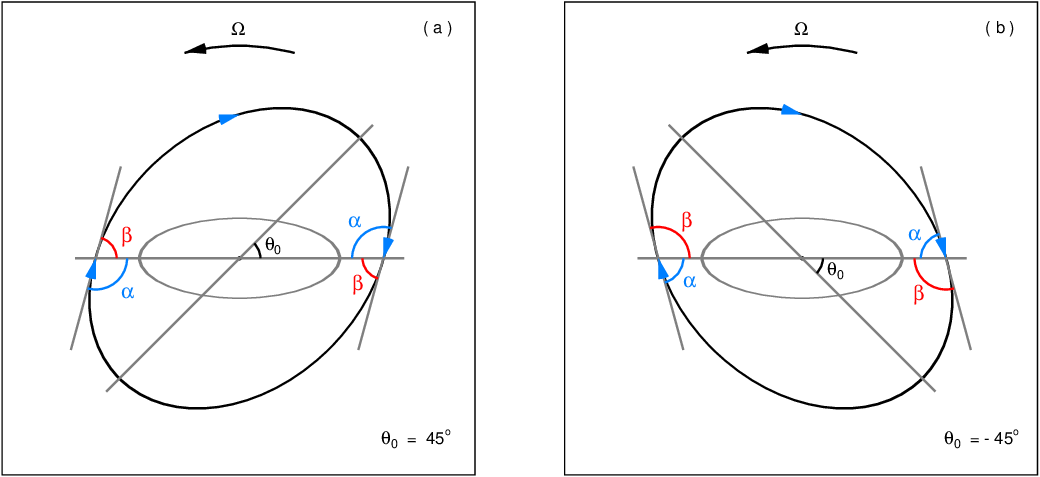}} \caption{Connection
between the changes in the angular momentum $L$ of a star and the
orientation of its orbit: $\theta_0>0$ (a) and $\theta_0<0$ (b). The
galaxy rotates counterclockwise. The star in the bar's reference
frame rotates clockwise. During one revolution around the bar, the
star loses and gains angular momentum $L$ twice, moving toward and
away from the galactic center, respectively. (a) For the orbital
orientation with the tilt  to the right, $\theta_0>0$, the star
approaches the ends of the bar at an obtuse angle $\alpha$ and moves
away at an acute angle $\beta$, $\alpha>\beta$. The bar first pulls
the star clockwise, causing a decrease in $L$, and then
counterclockwise, causing an increase in $L$. The star moves away
from the major axis of the bar at an acute angle and remains for some
time closer to the end of the bar, which causes it to acquire
additional angular momentum. (b) For an orbit with the tilt to the
left, $\theta_0<0$, the star approaches the ends of the bar at an
acute angle $\alpha$ and moves away at an obtuse angle $\beta$,
$\alpha<\beta$. In this case, it loses more angular momentum than it
gains.} \label{fig_tilt}
\end{figure*}

The direction of the orbital elongation remains constant only for
periodic orbits oriented along the major or minor axes of the bar.
Let's compare the periods of the radial $T_r$ and azimuthal
$T_\varphi$ oscillations of the star under consideration,
corresponding to the frequencies $\kappa$ and $m(\Omega-\Omega_b)$:

\begin{equation}
T_r=\frac{2\pi}{\kappa},
 \label{eq:t_r}
\end{equation}

\begin{equation}
T_\varphi=\Bigl|\frac{2\pi}{m(\Omega-\Omega_b)} \Bigl|.
 \label{eq:t_phi}
\end{equation}

Figure~\ref{fig_t_phi_t_r}a shows variations in the periods of the
radial $T_r$ (red circles) and azimuthal $T_\varphi$ (blue circles)
oscillations of the star under consideration. If $T_r=T_\varphi$,
then the direction of the orbital elongation remains constant. If the
period of the radial oscillations $T_r$ is slightly longer than the
period of the azimuthal oscillations $T_\varphi$, then the orbital
elongation direction must shift in the sense opposite that of
galactic rotation. Indeed, it is the radial oscillations of a star
that determine the direction of the orbital elongation. In the bar's
reference frame, a star located near the OLR moves in the sense
opposite that of  galactic rotation,  so maximum distances $R$ must
gradually shift in a direction opposite that of galactic rotation.
And, conversely, if $T_r$ is slightly less than $T_\varphi$, then the
orbital elongation direction  must shift in the sense of galactic
rotation.

Figure~\ref{fig_t_phi_t_r}b shows variations in the slow angle
$\overline{\theta_s}$ (red circles), calculated from
Eq.~\ref{eq:theta_s}, and the angle $\theta_0$, obtained with the use
of numerical simulations. Both angles characterize the orbital
orientation relative to the major axis of the bar, but the angle
$\theta_0$ is calculated using the dependence of the distance $R$ on
the azimuthal angle $\varphi$, while the angle $\overline{\theta_s}$
is calculated from Eq.~\ref{eq:theta_s} with a step of 10 Myr and
then averaged over the interval of one radial oscillation. It is seen
that the angles $\overline{\theta_s}$ and $\theta_0$ oscillate in
phase, which is caused by the choice of the initial phase of the
oscillations of the angle $\theta_s$ and will be discussed below.
Moreover, the amplitude of the $\overline{\theta_s}$  oscillation is
approximately two times greater than the amplitude of the $\theta_0$
variations. This difference in amplitudes is likely caused by the
fast component in the oscillations of the $\theta_s$ angle
(Fig.~\ref{fig_ang_act}a).

A comparison of Fig.~\ref{fig_t_phi_t_r}a and \ref{fig_t_phi_t_r}b
shows that the increase (decrease) of the angle $\theta_0$ occurs at
time intervals when $T_\varphi>T_r$ ($T_r>T_\varphi$). This increase
or decrease in the angle $\theta_0$ is the result of the orbital
precession in the direction of the galaxy's rotation or in the
opposite direction, respectively.

Figure ~\ref{fig_tilt} illustrates the connection between the change
in the angular momentum $L$ of a star and its orbital orientation:
$\theta_0>0$ (a) and $\theta_0<0$ (b). The galaxy rotates
counterclockwise. The star in the bar's reference frame rotates
clockwise. During one revolution around the bar, the star loses and
gains angular momentum $L$ twice, moving closer to and further away
from the galactic center. When the orbit is tilted to the right
($\theta_0>0$, Figure ~\ref{fig_tilt}a), the star approaches the ends
of the bar at an obtuse angle $\alpha$ and moves away at an acute
angle $\beta$, $\alpha>\beta$. The gravitational force created by the
bar's ends first pulls the star clockwise, causing a decrease in $L$,
and then counterclockwise, causing an increase in $L$. The star,
moving away from the major axis of the bar at an acute angle, remains
for some time closer to the end of the bar, and therefore gains more
$L$ than it loses as it approaches the major axis of the bar. When
the orbit is tilted to the left, ($\theta_0<0$,
Fig.~\ref{fig_tilt}b), the star approaches the ends of the bar at an
acute angle $\alpha$ and moves away at an obtuse angle $\beta$,
$\alpha<\beta$. In this case, it loses more angular momentum than it
gains as it approaches the bar's ends.

Note that oscillations of the angular momentum $L$ of a star  are
accompanied by oscillations of the total energy $E$, so the Jacobi
integral $E_J$ (Eq.~\ref{eq:E_J}) retains its value.

An increase in the distance $R_g$  leads to a decrease in the average
angular velocity $\overline{\Omega}$ of the star's rotation
(Fig.~\ref{fig_r_omega_p}ab) and  the epicyclic frequency $\kappa$
(Eq.~\ref{eq:kappa}). The decrease in the values of
$\overline{\Omega}$ and $\kappa$ with increasing distance $R_g$ is a
general characteristic of differential galactic rotation.
Consequently, with increasing $R_g$, the resonance frequency
$\Omega_s$ (Eq.~\ref{eq:omega_s2}) must decrease. Indeed, a
comparison of Fig.~\ref{fig_r_omega_p}a and Fig.~\ref{fig_omega_s}a
shows that the variables $R_g$ and $\Omega_s$ oscillate out of phase.

The specific feature of stellar motion near the OLR is that an
increase in $R_g$ leads to a decrease in both quantities
$\overline{\Omega}$ and $\kappa$, but affects the periods of the
radial and azimuthal oscillations differently: $T_r$ increases, while
$T_\varphi$ decreases (Fig.~\ref{fig_t_phi_t_r},
Eq.~\ref{eq:t_r},~\ref{eq:t_phi}). This distinction is caused by the
fact that near the OLR, the difference $(\Omega - \Omega_b)<0$.

An increase in the average angular momentum, $L$, of the star is
accompanied by an increase in the distance, $R_g$, and corresponds to
positive values of the angle $\theta_0$ (Fig.~\ref{fig_ang_act}b,
\ref{fig_r_omega_p}a, \ref{fig_t_phi_t_r}b, \ref{fig_tilt}). An
increase in $L$ and $R_g$ leads to a decrease in the values of
$\Omega$ and $\kappa$ which, in turn, causes a decrease in
$T_\varphi$ and an increase in $T_r$. At the moment when
$T_r=T_\varphi$, the angle $\theta_0$ reaches its maximum value. If
this value lies in the range $0<\theta_0<90^\circ$ then $L$ and $R_g$
continue to increase, but the orbital precession changes its
direction and the angle $\theta_0$ begins to  decrease.

As soon as the orbit of the star becomes elongated along the major
axis of the bar, the situation changes. Although $\theta_0$ continues
to decrease ($T_r>T_\varphi$), the star begins to lose angular
momentum $L$ (Fig.~\ref{fig_tilt}b). This causes a decrease in $R_g$
and increase in  $\Omega$ and $\kappa$, which leads to an increase in
$T_\varphi$ and a decrease in $T_r$. At the moment when
$T_r=T_\varphi$, the angle $\theta_0$ reaches its minimum value. If
this value lies in the range $-90^\circ<\theta_0<0$, then $L$ and
$R_g$ continue to decrease but the orbital precession reverses its
sense and the angle $\theta_0$ begins to increase.

When the orbit is oriented along the major axis of the bar
($\theta_0=0$), $L$ and $R_g$ reach their extreme values, with the
maxima corresponding to moments when $\theta_0$ decreases, and the
minima --- to moments when $\theta_0$ increases
(Fig.~\ref{fig_ang_act}ab, ~\ref{fig_r_omega_p}a).

Note that the angle $\theta_R$ (Eq.~\ref{eq:theta_R}) and,
consequently, the angle $\theta_s$ (Eq.~\ref{eq:theta_s}) are
determined up to an additive shift of $\pm\pi$. In our case, the
initial phase of oscillations of the angle $\theta_s$ must be shifted
by $\pi$, i.~e., the angle $\theta_s$ must be replaced by
$\theta_s+\pi$. This choice of the initial phase of the $\theta_s$
oscillations allows us to provide the correct signs of the variables
in Eq.~\ref{eq:dot_J_s}, namely, an increase in $J_s$ for $c_{12}<0$
must correspond to negative values of $\theta_s$.

Near the OLR of the bar, there are a large number of orbits ($\sim
2/3$) which are not captured by libration and their orbital
precession goes in only one direction \citep{melnik2024}.  With high
accuracy, the direction of orbital precession is determined by the
sign of the resonance frequency $\Omega_s$
(Eq.~\ref{eq:dot_theta_s}). The value of the additional term in this
equation, $\partial{c_{lm}}/\partial{J_s}$, calculated for a typical
star (Section 4.1), does not exceed 0.1 km s$^{-1}$ kpc$^{-1}$. If
$\Omega_s$ takes both positive and negative values, then the orbit
changes the direction of precession, i.~e., librates. If $\Omega_s$
takes only positive values ($T_r<T_\varphi$) or only negative values
($T_r>T_\varphi$), then the orbit precesses in only one direction.

Orbits captured by  libration  near the OLR are characterized by a
larger amplitude of the slow changes in the angular momentum $L$, and
consequently, in the average distance $\overline{R}$. Changes in
$\overline{L}$ and $\overline{R}$  must cause a change in the
relationship between the periods of the radial and azimuthal
oscillations. In general, the larger the amplitude of the slow
changes in $L$, the greater the range of variations in the distance
$\overline{R}$, and the higher the probability that the ratio between
the periods $T_r$ and $T_\varphi$ changes, and therefore, orbital
precession  changes its sense.

\subsection{4.3 Dependence of the orbital libration frequency  on the parameters of
the bar and other galactic components}

Let us derive a more general expression for the libration frequency
of  the orbital elongation direction of stars near the  OLR. We will
consider stars with an average azimuthal velocity close to that of
the rotation curve, $V_c$. It is convenient to compare the
oscillation frequency of the orbital elongation direction, $\omega$,
with the epicyclic frequency $\kappa$ at the distance of $R_{\rm
OLR}$.

Let us  make the following assumptions. In Eq.~\ref{eq:omega^2}, we
replace the factor $\sin \theta_s/\theta_s$, which depends on the
angle $\theta_s$, with its average value, $c_t$.
Figure~\ref{fig_ang_act}a shows that the angle $\theta_s$ lies in the
range $\sim [-90,90^\circ]$. Then the coefficient $c_t$ is determined
by the  expression

\begin{equation}
c_t=\frac{1}{\pi} \int_{-\pi/2}^{\pi/2} \frac{\sin \theta}{\theta}
d\theta
 \label{eq:c_t}
\end{equation}

\noindent and is equal to $c_t=0.8727$. Therefore, the average value
of the frequency $\omega$ is determined by only two variables,
$\partial{\Omega_s}/\partial{J_s}$ and $c_{12}$:

\begin{equation}
\omega^2= -c_t \frac{\partial{\Omega_s}}{\partial{J_s}} c_{12}.
 \label{eq:omega^2_new}
\end{equation}

\noindent Furthermore, we will assume that the Galactic rotation
curve is flat at the distance of $R_{\rm OLR}$. Then the frequency
$\kappa$ is related to the average angular rotation frequency of a
star $\Omega$ by the relation:

\begin{equation}
\kappa \approx \sqrt{2}\Omega.
 \label{eq:kappa_omega}
\end{equation}

\noindent Using Eqs.~\ref{eq:dir_omega_s}, \ref{eq:kappa_omega} and
taking into account that the angular momentum of the star must have a
value close to $L=\Omega R^2$, we can express the derivative
$\partial{\Omega_s}/\partial{J_s}$ as:

\begin{equation}
\frac{\partial{\Omega_s}}{\partial{J_s}}=\frac{1}{R^2}
(4+\frac{6}{\sqrt{2}}) \approx \frac{8.24}{R^2}.
 \label{eq:dir_omega_s_new}
\end{equation}

\noindent The Fourier coefficient $c_{12}$ (Eq.~\ref{eq:c_jk})  can
be written in the following way:

\begin{equation}
c_{12}=- (\gamma + \frac{1}{2}) \, \alpha_b \, \overline{p} \, R \,
\frac{d \Phi_b}{d{R}}, \label{eq:c_12_new}
\end{equation}

\noindent where we used the fact that the average (over the azimuthal
angle) bar potential, $\Phi_b(R)$, can be approximated by the
expression $\Phi_b(R)=K a/R$ (Eq.~\ref{eq:a/R}).

Let us introduce the coefficient $f_b$, which connects the average
radial force generated by the bar at the distance  $R_{\rm OLR}$ with
the average radial force generated by the entire galaxy (by all its
constituents) at the same distance:

\begin{equation}
 f_b=  \frac{d{\Phi_b}/d{R}}{<\partial{\Phi}/\partial{R}>}.
\label{eq:epsilon_b}
\end{equation}

\noindent Note that the average radial force, created by the galaxy
at a distance $R$, is related to the velocity of rotation curve by
the following relation:

\begin{equation}
V_c^2=<\frac{\partial{\Phi}}{\partial{R}}> R. \label{eq:v_c_2}
\end{equation}

Considering Eq.~\ref{eq:c_t} -- \ref{eq:v_c_2}, as well as the ratios
$\Omega \approx V_c/R$ and $\gamma=1.40$, we can write the expression
for the orbital oscillation frequency, $\omega$, as follows:

\begin{equation}
\omega^2  \approx 6.83 \alpha_b f_b \overline{p} \,\kappa^2.
 \label{eq:omega^2_f}
\end{equation}

In our model, the average epicyclic frequency $\kappa$ at the OLR
distance, obtained without taking into account asymmetric drift, is
$\kappa=45.50$ km s$^{-1}$ kpc$^{-1}$. The coefficients $\alpha_b$
and $f_b$ at the OLR distance are $\alpha_b=0.0562$ and $f_b=0.1520$.
The average orbital elongation $\overline{p}$ is the most difficult
to estimate, but here we will use the average value obtained for a
typical star over the time interval 0--6 Gyr, $\overline{p}=0.093$
(Section 4.1, Fig.~\ref{fig_r_omega_p}c). Substituting these values
into Eq.~\ref{eq:omega^2_f}, we obtain $\omega=3.4$ Gyr$^{-1}$, which
corresponds to the period of $P=1.8$ Gyr.

The estimate of the orbital oscillation period  of $P=1.8$ Gyr is
only 10\% smaller than the value of $P=2.0$ Gyr obtained in Section
4.1, which may be related to the numerous simplifying assumptions
used in this section. One of these concerns asymmetric drift (see
also Section 4.1), which we neglected. If we take into account that
stars at the OLR distance rotate with the average velocity of
$\overline{V_\varphi}=V_c-7$ km s$^{-1}$, then the value of $\kappa$
decreases to $\kappa=43.50$ km s$^{-1}$ kpc$^{-1}$
(Eq.~\ref{eq:kappa}), and the oscillation period of librating orbits
increases to 1.9 Gyr.

Note that the coefficients $\alpha_b$ and $f_b$ characterize
different properties of the bar. The coefficient $\alpha_b$, equal to
the ratio of the amplitude of the radial force generated by the bar
to the average radial force created by the bar, mainly characterizes
the bar's shape: the more elongated the bar, the higher the value of
$\alpha_b$, and if the bar degenerates into an oblate spheroid, then
$\alpha_b=0$. As for the coefficient $f_b$, equal to the ratio of the
average radial force generated by the bar to the radial force created
by the entire galaxy, it rather characterizes the ratio of the bar's
mass to the mass of the disk and  inner part of the halo.

The oscillation frequency of librating orbits $\omega$ strongly
depends on the epicyclic frequency $\kappa$ at the distance of the
OLR (Eq.~\ref{eq:omega^2_f}), which, in turn, depends on the angular
velocity of the disk rotation, $\kappa \sim \Omega$
(Eq.~\ref{eq:kappa_omega}).

Furthermore, the frequency of oscillations of  librating orbits
$\omega$ depends on the position of the OLR in the galaxy. The slower
the bar rotates, the farther from the center the OLR radius is
located, and the smaller the value of $\kappa$ at this radius.
Consequently, the smaller $\Omega_b$, the lower the frequency
$\omega$ and the longer the period of oscillations of librating
orbits.

The coefficients $\alpha_b$ and $f_b$ can be combined into a single
coefficient, $q_R$, equal to the ratio of the amplitude of the radial
force generated by the bar at the OLR distance to the average radial
force created by the entire galaxy at this distance:

\begin{equation}
q_R= \alpha_b f_b.
 \label{eq:q_R}
\end{equation}

\noindent The coefficient $q_R$ is analogous to the coefficient $q_t$
(Section 2), which characterizes the ratio of the amplitude of the
azimuthal force to the average radial force created by the entire
galaxy. In our model, $q_R=0.0084$.

Then, Eq.~\ref{eq:omega^2_f} can be rewritten as:

\begin{equation}
\omega^2  \approx 6.83 q_R \overline{p} \,\kappa^2.
 \label{eq:omega^2_final}
\end{equation}

Thus, the frequency of oscillations of librating orbits near the OLR
of the bar is determined by the parameters of the bar (shape,
relative mass, angular rotation velocity), as well as the combined
mass of the bar, disk, and inner halo (Eq.~\ref{eq:omega^2_final}).

\section{5. Conclusions}

We studied the orbit of a typical star trapped by libration near the
OLR of the Galactic bar (Fig.~\ref{fig_orbit}) in angle-action
variables. The variations in the slow angle $\theta_s$ and the
corresponding action $J_s$ ($J_s=L/m$) have both  fast and  slow
components (Fig.~\ref{fig_ang_act}). The slow angle characterizes the
direction of the orbital elongation while the fast angle determines
the phase of the motion of a star along its orbit. The slow
variations in $\theta_s$ and $J_s$, obtained from numerical modeling
of the orbit of  the star under consideration, have the period of
$P=2.0$ Gyr.

Using the analytical theory described in \citep{weinberg1994,
binney2008, monari2016, monari2017}, we calculated the values of the
frequency $\omega$ of oscillations of the slow angle $\theta_s$,
averaged over the time interval of one radial oscillation of the
chosen star (Eq.~\ref{eq:omega^2}). The Galactocentric distance to
the  center of epicycle $R_g$ was taken to be equal to the average
distance to the star $\overline{R}$, calculated over the interval of
one radial oscillation (Eq.~\ref{eq:r_g}). It turned out that the
frequency of libration $\omega$ changes with the distance $R_g$, but
the value of $\omega$, averaged over the time interval $t=0$--6 Gyr,
is equal to $P=2.0$ Gyr (Fig.~\ref{fig_w_per}). Thus, the periods of
libration derived from theory and numerical simulations are in good
agreement.

We have shown that when the orbit is tilted in the direction of the
galactic rotation relative to the minor axis of the bar
($\theta_0<0$), the angular momentum of the star decreases on
average, while when the orbit is tilted in the opposite direction
($\theta_0>0$), it increases. An increase (decrease) in the angular
momentum $\overline{L}$ causes an increase (decrease) in the average
distance $\overline{R}$, which can lead to a change in the
relationship between the periods of the radial and azimuthal
oscillations of a star: for $T_r>T_\varphi$, the angle $\theta_0$,
which characterizes the orbital orientation, decreases, while for
$T_\varphi>T_r$, it increases.  Thus, the change in the average
radius $\overline{R}$ is regulated by  the change in orbital tilt.

By making a number of simplifications, we obtained a more general
formula for the frequency of oscillations in the orbital elongation
direction, $\omega$, of orbits captured by libration near the OLR of
the bar (Eq.~\ref{eq:omega^2_final}). The ratio of the frequency
$\omega$ to the epicyclic frequency at the OLR distance,
$\omega/\kappa$, is determined by the parameters $q_R$ and
$\overline{p}$. The coefficient $q_R$ is the ratio of the amplitude
of the radial force, generated by the bar, to the average radial
force created by the entire galaxy and characterizes the relative
value of disturbances generated by the bar at the OLR distance. The
coefficient $\overline{p}$ characterizes the average elongation of
orbits captured by libration near the OLR. The  period of libration
of orbits near the OLR of the Galactic bar, obtained from
Eq.~\ref{eq:omega^2_final}, is 1.8 Gyr, which is in good agreement
with the data of modeling.

\section*{Acknowledgments}

{\footnotesize We thank the anonymous referee for helpful remarks and
suggestions.   This work was carried out using data from the European
Space Agency (ESA) mission {\it Gaia} (https://
www.cosmos.esa.int/gaia), processed by the Data Processing and
Analysis Consortium (DPAC, https://www.cosmos.esa.int/web/gaia
/dpac/consortium) {\it Gaia}. DPAC support was provided by national
institutions, in particular institutions participating in the
multilateral agreement {\it Gaia}.

\section*{Funding}

The study was conducted under the state assignment of Lomonosov
Moscow State University.

\section*{Conflict of Interest}

The authors of this work declare that they have no conflict of
interest.

\end{document}